\documentclass[
reprint,
amsmath,amssymb,
aps,
prl,
floatfix
]{revtex4-2}

\usepackage{graphicx}
\usepackage{dcolumn}
\usepackage{bm, bbm}
\usepackage[colorlinks, allcolors=blue]{hyperref} 
\def\equationautorefname#1#2\null{Eq.#1(#2\null)}

\usepackage{microtype}
\usepackage{tabularray}
\UseTblrLibrary{booktabs}

\usepackage{float}
\usepackage[dvipsnames]{xcolor}
\usepackage[caption=false]{subfig}
\newcommand{\ket}[1]{|#1\rangle}
\newcommand{\bra}[1]{\langle #1|}

\begin{document}

\preprint{APS/123-QED}

\title{Taming the Bloch-Redfield equation: Recovering an accurate Lindblad equation for general open quantum systems}

\author{Diego Fernández de la Pradilla}
    \email{diego.fernandez@uam.es}
    \affiliation{Departamento de Física Teórica de la Materia Condensada and Condensed Matter Physics Center (IFIMAC), Universidad Autónoma de Madrid, E-28049 Madrid, Spain}
\author{Esteban Moreno}%
    \email{esteban.moreno@uam.es}
    \affiliation{Departamento de Física Teórica de la Materia Condensada and Condensed Matter Physics Center (IFIMAC), Universidad Autónoma de Madrid, E-28049 Madrid, Spain}
\author{Johannes Feist}%
    \email{johannes.feist@uam.es}
    \affiliation{Departamento de Física Teórica de la Materia Condensada and Condensed Matter Physics Center (IFIMAC), Universidad Autónoma de Madrid, E-28049 Madrid, Spain}
\date{\today}

\begin{abstract}
    Master equations play a pivotal role in investigating open quantum systems. In particular, the Bloch-Redfield equation stands out due to its relation to a concrete physical environment. However, without further approximations it does not lead to a Lindblad master equation that guarantees that the density matrix stays completely positive, which has raised some concerns regarding its use. This study builds on previous efforts to transform the Bloch-Redfield framework into a mathematically robust Lindblad equation, while fully preserving the effects that are lost within the secular approximation that is commonly used to guarantee positivity. These previous approaches introduce two potential deficiencies: the environment-induced energy shift can be non-Hermitian and some decay rates can be negative, violating the assumptions of Lindblad's theorem. Here, we propose and evaluate straightforward solutions to both problems. Our approach offers an effective and general procedure for obtaining a Lindblad equation, derived from a concrete physical environment, while mitigating the unphysical dynamics present in the Bloch-Redfield equation.
\end{abstract}

\maketitle

\section{Introduction}

Inevitably, any quantum mechanical system interacts with its surroundings~\cite{schlosshauer_decoherence_2007}. Therefore, it is not possible to fully understand the dynamics of a system without properly accounting for the environment. In practice, direct approaches to calculate the evolution with the Schrödinger equation are far from feasible, due to the large (even infinite) number of degrees of freedom in the environment. Consequently, it becomes imperative to develop approximate methods that identify and preserve the relevant information. 
Typically, one's primary interest is the system, and the system-environment coupling is weak enough that only negligible correlations appear between the two. In this context, the framework of open quantum systems~\cite{breuer_openquantsys_2002} provides a description of the system's time evolution through so-called quantum master equations (QMEs). By incorporating the effect of the environment perturbatively, QMEs have garnered use in most fields of physics, from condensed matter physics and quantum optics~\cite{harbola_electrontransport_2006,hartle_etransport-qdots_2013,carmichael_open_1993,lambropoulos_quantum-optics_2000} to high-energy physics and cosmology~\cite{diosi_qedme_1990,colas_cosmologicalme_2022}. However, despite their widespread use, QMEs are perturbative in nature and rely on approximations whose physical meaning and mathematical accuracy have to be considered. More specifically, different QMEs can depend on varying degrees of approximation, which can in turn invalidate one in favor of another.

Lindblad's theorem shows that any Markovian QME corresponding to a trace-preserving and completely positive map can be written as a Gorini-Kossakowski-Sudarshan-Lindblad master equation~\cite{lindblad_quantumsemigroups_1976,manzano_lindblad_2020,breuer_openquantsys_2002}, from here on out called Lindblad master equation (LME) for conciseness. It is given by
\begin{equation} \label{eq.LME}
    \dot{\rho} = -\frac{i}{\hbar}[H+\hbar\Delta,\rho] + \sum_{ij} \frac{\gamma_{ij}}{2} \Big( -\{A_i^\dagger A_j,\rho\} + 2A_j\rho A_i^\dagger  \Big)
\end{equation}
and guarantees that the density matrix stays physical at all times, i.e., that it is Hermitian, has trace equal to one, and only contains non-negative diagonal entries. Here, \(\rho\) is the system's density matrix, \(H\) is the bare system Hamiltonian, \(\hbar\Delta\) is the (Hermitian) Hamiltonian induced by the interaction with the environment, \(A_i\) are the collapse operators and \(\gamma_{ij}\) is a positive semidefinite matrix known as the Kossakowski matrix. The first term on the right-hand side describes the unitary evolution of the system, while the second term describes the dissipation induced by the environment. The LME is commonly written in a slightly simpler form obtained by transforming to the eigenbasis of the Kossakowski matrix \(\gamma_{ij}\), which makes the last term diagonal in \(i\) and \(j\). While many works use ad-hoc LMEs with dissipation rates and operators chosen manually, this can lead to unphysical effects, especially if the system consists of coupled subsystems~\cite{carmichael_master_1973,gambetta_dissipation_2011,delpino_quantum_2015,stassi_output_2016,settineri_dissipation_2018,lednev_lindblad_2023}. A more systematic approach consists in starting from a microscopic Hamiltonian including both system and bath, and then tracing out the environmental degrees of freedom. Unfortunately, this procedure does not yield an LME in general, but another QME known as the Bloch-Redfield equation (BRE)~\cite{breuer_openquantsys_2002}, which does not guarantee positivity of the populations and can thus lead to unphysical density matrices. To finally obtain an LME, the traditional approach is to perform the so-called secular approximation on the BRE, in which terms that oscillate (within the interaction picture) are assumed to average out and are removed. However, it has been shown that this step is not always well-justified or accurate~\cite{jeske_bloch-redfield_2015,eastham_bath-induced_2016,hartmann_embracing-nonpositivity_2020}. 

Indeed, the secularization procedure completely eliminates the couplings between populations and coherences and can therefore miss significant effects. This is the case if, for instance, the energy spectrum has levels that are close to each other compared to the energy scale associated to the coupling to the environment. With the secular approximation, the resulting LME disregards important interference effects between these states, while the BRE includes the relevant couplings. Its mathematical flaws, however, have motivated the development of improved procedures that yield an LME and overcome the limitations of the secular approximation~\cite{farina_positivityRedfield_2019,davidovic_completelypositive_2020,mccauley_accurate-master_2020,nathan_universallindblad_2020,dabbruzzo_timedepRedfield_2023}. In general, it has been shown that any alteration to the BRE will incur in certain problems regarding thermodynamic properties or local conservation laws~\cite{tupkary_fundlimitation_2022,tupkary_localconservation_2023}. Fixing such issues requires, therefore, alternative procedures such as those explored in~\cite{becker_lindbladapprox_2021,becker_canonicallyconsistent_2023,schnell_globallocal_2023}, to go to higher orders in the coupling to the bath or more accurate alternatives to the BRE. We do not consider here such problems, and focus on the simpler goal of retrieving an effective LME from the BRE. Among the available methods, the approach proposed in~\cite{mccauley_accurate-master_2020} is notable for its simplicity and effectiveness. There, the authors propose a prescription that works well for simple system-environment interaction Hamiltonians within the rotating wave approximation. We further extended those results in~\cite{dfpv_vacuum-field-induced_2023} by incorporating counter-rotating terms into more complex system-environment interactions. Still, the master equation used in~\cite{dfpv_vacuum-field-induced_2023} can lead to two undesired possibilities: (i) non-Hermitian, off-diagonal couplings, depending on the interplay between the energy structure of the system and the resonances of the bath, and (ii) a non-positive Kossakowski matrix if there are cross-correlations between bath operators. Either option invalidates the requirements for an LME, and both are a direct consequence of the application of the methodology of~\cite{mccauley_accurate-master_2020} to complex systems and environments. Notwithstanding, under conditions of geometrical symmetry and with a careful treatment of the system level structure, these difficulties can be overcome. Indeed, we took this approach in~\cite{dfpv_vacuum-field-induced_2023} to reveal strong Casimir-Polder-induced state mixing and counterintuitive state protection in a system involving a hydrogen atom and a dielectric nanoparticle. 

The goal of the present article is to present a way to deal with arbitrary setups by modifying the prescription of~\cite{mccauley_accurate-master_2020}. We introduce and compare several alternatives that resolve the issues of non-Hermitian couplings and non-positive semidefiniteness (NPS) of the Kossakowski matrix, while maintaining an accurate description of the dynamics, turning the BRE into a general and precise LME. On one hand, we find that the non-Hermiticity of the energy shift can be resolved by using the arithmetic instead of the geometric mean in the symmetrization procedure for the environment-induced energy shift. On the other hand, the NPS issue can be resolved effectively by discarding the negative eigenvalues of the Kossakowski matrix. We explain the details of the procedures and justify them with model examples that capture the essence of each problem. The Python library QuTiP~\cite{qutip} has been used throughout this work for the simulations that illustrate and support our claims. The article is organized as follows: In \autoref{sec.theory}, we discuss the BRE and its deficiencies explicitly, along with the explanation of the methodology developed in~\cite{mccauley_accurate-master_2020} to turn it into an LME\@. The non-Hermiticity issue is addressed in \autoref{sec.nonHermiticity}, where two possibilities are compared and, eventually, one is identified as the best. The analysis of the NPS of the Kossakowski matrix is presented in \autoref{sec.nonpositivity}. Finally, we summarize the results of our exploration in \autoref{sec.conclusion}.

\section{Theory} \label{sec.theory}

In this section, we write the BRE and elucidate how the method proposed in~\cite{mccauley_accurate-master_2020} transforms it into an LME, subject to certain conditions. Deviation from these conditions implies that the resulting equation is not an LME, which motivates the development of the extensions developed in this article.

\subsection{Bloch-Redfield master equation} \label{subsec.bre}

\begin{figure*}[ht]
    \centering
    \subfloat{\label{subfig.BRE-00-map}}
    \subfloat{\label{subfig.BRE-11-map}}
    \subfloat{\label{subfig.BRE-22-map}}
    \subfloat{\label{subfig.BRE-00-cut}}
    \subfloat{\label{subfig.BRE-11-cut}}
    \subfloat{\label{subfig.BRE-22-cut}}
    \includegraphics[width=\textwidth]{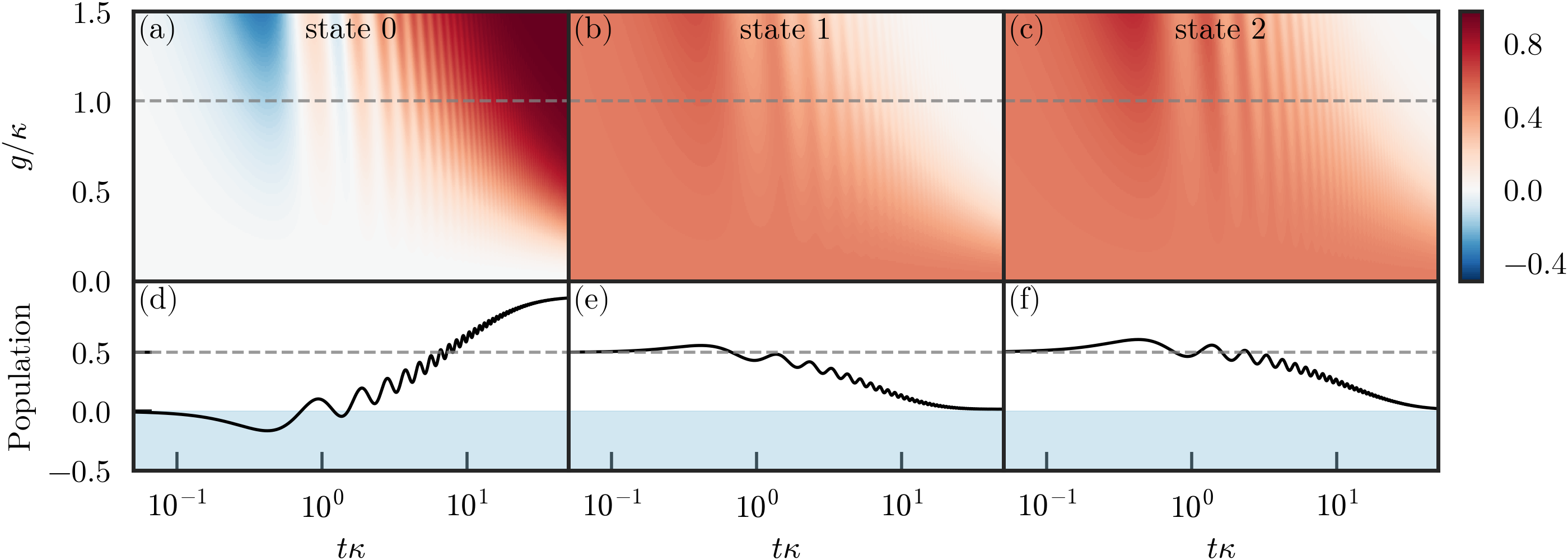}
    \caption{Breakdown of population positivity with the BRE. Top panels: Population of the states of a three-level system, as a function of the coupling strength, \(g/\kappa\), and time, \(t\kappa\). Blue and red indicate negative and positive population, respectively. Bottom panels: Same populations at \(g=\kappa\) (dashed lines in the top panels). The blue region represents negative, unphysical population, and the dashed line offers a visual aid to see how the population of the states \(\ket{1}\) and \(\ket{2}\) add up to more than one. \label{fig.BRE-negative}}
\end{figure*}

The BRE serves as starting point in numerous derivations of LMEs, due to its physically motivated derivation. We write it for a general system-bath interaction \(H_\mathrm{sb}=\sum_\alpha A_\alpha B_\alpha\), where \(A_\alpha\) and \(B_\alpha\) represent system and bath operators, respectively, and at zero temperature:
\begin{align}\nonumber
    \dot{\rho} = -\frac{i}{\hbar}[H&,\rho]  \\ \nonumber 
     +\sum_{ij}\bigg\{ 
        &-i\Big(\Lambda_{ij}(\omega_{j})\sigma_i^\dagger\sigma_j\rho - \rho\Lambda_{ij}(\omega_{i})\sigma_i^\dagger\sigma_j \Big)\\ \nonumber
        &+i\Big(\Lambda_{ij}(\omega_j) - \Lambda_{ij}(\omega_i) \Big) \sigma_j\rho\sigma_i^\dagger\\ \nonumber
        &-\frac{1}{2}\Big(\Gamma_{ij}(\omega_{j})\sigma_i^\dagger\sigma_j\rho + \rho\Gamma_{ij}(\omega_{i})\sigma_i^\dagger\sigma_j \Big)\\ \label{eq.BRE}
        &+\frac{1}{2}\Big(\Gamma_{ij}(\omega_{j}) + \Gamma_{ij}(\omega_{i})\Big)\sigma_j\rho\sigma_i^\dagger \bigg\}.
\end{align}
In the above equation, \(\rho\) and \(H\) represent the system's density matrix and Hamiltonian, the indices \(i\) and \(j\) are combined indices that denote transitions, with \(\sigma_j = \ket{n_j}\bra{m_j}\) being the \(j\)th transition operator between eigenstates \(\ket{n_j}\) and \(\ket{m_j}\) of \(H\), \(\omega_{j} = (E_{m_j}-E_{n_j})/\hbar\) is the \(j\)th transition frequency, and the functions \(\Gamma_{ij}\) and \(\Lambda_{ij}\) are defined as
\begin{subequations}
    \begin{align}
        \Gamma_{ij}(\omega) &= \sum_{\alpha\beta}(A_\alpha)_{i}^* \gamma_{\alpha\beta}(\omega)(A_\beta)_{j},\\
        \Lambda_{ij}(\omega) &= \sum_{\alpha\beta}(A_\alpha)_{i}^* \lambda_{\alpha\beta}(\omega)(A_\beta)_{j}.
    \end{align}
\end{subequations}
Here, \((A_\beta)_{j} = \mathrm{Tr}\{\sigma_j^\dagger A_\beta\}\) is the transition matrix element associated with the \(j\)th transition, and the real functions \(\gamma_{\alpha\beta}\) and \(\lambda_{\alpha\beta}\) are given by
\begin{subequations}
    \begin{equation} \label{eq.gamma}
        \frac{\gamma_{\alpha\beta}(\omega)}{2} + i \lambda_{\alpha\beta} = \int_{0}^\infty \mathrm{d}\tau\ \langle B_\alpha(\tau)B_\beta(0) \rangle e^{i\omega\tau},
    \end{equation}
    where the imaginary and real parts are related through a Kramers-Kronig-type relation:
    \begin{equation}
        \lambda_{\alpha\beta}(\omega) = \frac{1}{2\pi}\mathrm{P}\int \mathrm{d}\omega'\ \frac{\gamma_{\alpha\beta}(\omega')}{\omega-\omega'}.
    \end{equation}
\end{subequations}
The P denotes the principal value and \(\boldsymbol{\gamma}\) is a non-diagonal and positive definite tensor-valued function. It is related to the so-called spectral density \(\mathbf{J}\) through \(\boldsymbol{\gamma} = 2\pi\mathbf{J}\), which characterizes the coupling between the system and the bath. The limitation to zero temperature comes from the assumption in the derivation that the bath is approximately in the vacuum, or in a thermal state such that \(k_B T \ll \min_{i}(\omega_i)\). Not only is this a common circumstance within, e.g., quantum optical systems, but such an approximation is also not strictly necessary, and is done here only for simplicity.

Roughly speaking, \(\Lambda_{ij}\) and \(\Gamma_{ij}\) can loosely be associated with the environment-induced energy shifts and decay rates, respectively. However, this distinction is not completely clear within the BRE itself, as the most natural interpretation depends on the specific arrangement of the equation. For instance, in~\autoref{eq.BRE}, we would recover an equation of Lindblad-form if the following conditions were met: (i) the second line should become a commutator, (ii) the third line should disappear, (iii) the fourth line should become an anticommutator, and (iv) the last line should become the ``refilling'' or ``quantum jump'' term, with the same prefactor as the previous line. Then, the second line would straightforwardly become the energy correction, and the last two lines would embody Lindblad-like dissipators, with clearly distinct roles for \(\Lambda\) and \(\Gamma\). These requirements are not directly satisfied in \autoref{eq.BRE} due to \(\Lambda_{ij}\) and \(\Gamma_{ij}\) being evaluated at two different frequencies in each line, \(\omega_j\) and \(\omega_i\), but the approach followed in~\cite{mccauley_accurate-master_2020} achieves conditions (i) through (iv) by means of an additional, well-justified approximation. In this paper we explore the same procedure. Nevertheless, it should be noted that the conditions above are not the only way to transform the BRE into a Lindblad-like equation. In fact, with the following commutator identity, 
\begin{equation}
    aO_1O_2+bO_2O_1=\frac{a+b}{2}\{O_1,O_2\} + \frac{a-b}{2} [O_1,O_2],
\end{equation}
\autoref{eq.BRE} can be reformulated as
\begin{align}\nonumber
    \dot{\rho} &= -\frac{i}{\hbar}[H,\rho] \\ \nonumber
     & +\sum_{ij}\bigg\{ -\frac{K^+_{ij}(\omega_{j}) - K^-_{ij}(\omega_{i})}{2}[\sigma_i^\dagger\sigma_j,\rho]\\ \label{eq.BRE2}
     &+\frac{K^+_{ij}(\omega_j) + K^-_{ij}(\omega_i)}{2} \Big(- \{\sigma_i^\dagger\sigma_j, \rho\} + 2\sigma_j\rho\sigma_i^\dagger\Big) 
     \bigg\},
\end{align}
where \(K^{\pm}_{ij}(\omega_j) = \Gamma_{ij}(\omega_j)/2 \pm i\Lambda_{ij}(\omega_j)\). In this form, the BRE looks like an LME, with both \(\Lambda\) and \(\Gamma\) appearing in the energy correction term and in the dissipators, blurring their individual role. Note that even written as \autoref{eq.BRE2}, the BRE is not an LME, because the corresponding Kossakowski matrix is not positive semidefinite. 

\begin{figure}[htbp]
    \centering
    \includegraphics[scale=1]{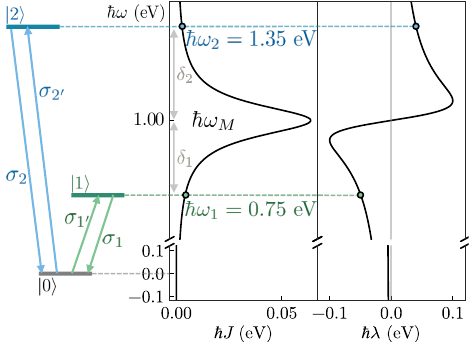}
    \caption{Side-by-side comparison of the system's level structure, \(J(\omega)\) and \(\lambda(\omega)\). The parameters for the figure are \(\hbar\omega_M = 1\)~eV, \(\hbar\kappa = 0.1\)~eV, \(\hbar g = 0.1\)~eV.}
    \label{fig.system}
\end{figure}

We conclude this subsection on the BRE by showing in \autoref{fig.BRE-negative} that \autoref{eq.BRE} can actually predict non-positive density matrices. To produce the figure, we have simulated a simple system composed of a three-level system coupled to a reservoir with a Lorentzian spectral density, schematically depicted in \autoref{fig.system}. The system eigenstates are \(\ket{0}\), \(\ket{1}\) and \(\ket{2}\), with two allowed lowering transitions from the excited states \(\ket{1}\) and \(\ket{2}\) to \(\ket{0}\), characterized by transition frequencies \(\omega_1\) and \(\omega_2\), respectively. Let us denote the transition operators with non-zero matrix element as
\begin{subequations} \label{eq.notation1}
    \begin{align} 
        \sigma_1 &= \ket{0}\bra{1}, & \sigma_2 &= \ket{0}\bra{2}, \\
        \sigma_{1'} &= \sigma_1^\dagger = \ket{1}\bra{0}, & \sigma_{2'} &= \sigma_2^\dagger = \ket{2}\bra{0},
    \end{align}
\end{subequations}
With this notation, the three-level system is coupled to a reservoir of harmonic oscillators through 
\begin{equation} \label{eq.systembath}
    H_\mathrm{sb} = \sum_{i=1}^{2}d_i\left(\sigma_i + \sigma_i^\dagger\right)\int \mathrm{d}\omega\  \sqrt{J(\omega)} \left(b_\omega + b^\dagger_\omega\right),
\end{equation}
where \(J(\omega)\) is the spectral density and, for concreteness, we set all \(d_i\) equal to each other and absorb them in the coupling to the environment. For a Lorentzian bath, 
\begin{subequations} \label{eq.notation2}
    \begin{equation} \label{eq.lorentzian}
        J(\omega) = \frac{g^2}{\pi} \frac{\kappa/2}{(\omega-\omega_M)^2 + (\kappa/2)^2},
    \end{equation}
    where \(g\) measures the coupling strength, with the absorbed transition matrix elements, and \(\omega_M\) and \(\kappa\) are the position and width of the peak of the mode described by the Lorentzian. The corresponding integral \(\lambda\) is analytical and expressed as
    \begin{equation} \label{eq.integral}
        \lambda(\omega) = \mathrm{P}\int_\mathbb{R} \mathrm{d}\omega' \frac{J(\omega')}{\omega-\omega'} = g^2 \frac{\omega-\omega_M}{(\omega-\omega_M)^2 + (\kappa/2)^2}.
    \end{equation}
\end{subequations}

The parameters are set to \(\hbar \omega_M = 1\)~eV, \(\hbar \kappa = 0.1\)~eV, \(\hbar \omega_1 = 0.75\)~eV, \(\hbar \omega_2 = 1.35\)~eV, and the initial state to \(\ket{\psi(0)}=(\ket{1} - \ket{2})/\sqrt{2}\). The population dynamics are depicted in \autoref{fig.BRE-negative} as a function of the coupling strength and time. In the top panels, negative and positive values are represented in blue and red, respectively. The bottom panels are a cut along the dashed lines, at \(g=\kappa\). In the first column, \autoref{subfig.BRE-00-map} and \autoref{subfig.BRE-00-cut} reveal a very prominent negative population in the groundstate \(\ket{0}\), an unmistakable sign of non-physicality. The preservation of the trace under BRE evolution implies that the populations of \(\ket{1}\) and \(\ket{2}\) together must exceed 1. This compensation is evident in the slight initial bumps observed in \autoref{subfig.BRE-11-cut} and \autoref{subfig.BRE-22-cut}, both above the 0.5 line.

\subsection{Prescription to obtain an LME} \label{subsec.mccauley}

In this subsection, we explain the procedure devised in~\cite{mccauley_accurate-master_2020} to turn the BRE into a Lindblad-like equation. We begin by noting that there are two terms in every line in the sum of~\autoref{eq.BRE}, evaluated at \(\omega_j\) and \(\omega_i\), respectively. The asymmetry in the indices \(i\) and \(j\) is the reason why the conditions (i) to (iv) stated in the previous subsection are not fulfilled. To overcome this difficulty, it was proposed to replace \(\Gamma_{ij}(\omega_j)\) and \(\Gamma_{ij}(\omega_i)\) by their geometric mean, \(\tilde{\Gamma}_{ij} = \sqrt{\Gamma_{ij}(\omega_j)}\sqrt{\Gamma_{ij}(\omega_i)}\), and likewise for \(\Lambda_{ij}\). This adjustment, whose accuracy is justified below, removes the difference in the roles of \(\omega_j\) and \(\omega_i\). We designate this prescription as \(g\Lambda-g\Gamma\), as both magnitudes are replaced by geometric averages. To be more exact,~\cite{mccauley_accurate-master_2020} deals with systems with one system-bath interaction, while this explicit formulation, first presented in~\cite{dfpv_vacuum-field-induced_2023}, accounts for multiple interactions, including cross-correlations between bath operators. 

After applying the replacement of the geometric mean in \autoref{eq.BRE}, the master equation is
\begin{align}\nonumber
    \dot{\rho} =& -\frac{i}{\hbar}[H,\rho] -i \sum_{ij} \tilde{\Lambda}_{ij}[\sigma_i^\dagger\sigma_j,\rho] \\ \label{eq.mccauley}
    &+ \sum_{ij}\frac{\tilde{\Gamma}_{ij}}{2}\Big(-\{\sigma_i^\dagger\sigma_j,\rho\} + 2\sigma_j\rho\sigma_i^\dagger\Big).
\end{align}
The justification for the replacement lies in the consideration of a slowly varying spectral density, where \(\mathbf{J}(\omega + \Lambda)\simeq \mathbf{J}(\omega)\) and \(\mathbf{J}(\omega + \Gamma)\simeq \mathbf{J}(\omega)\). Under these conditions, every term in \autoref{eq.BRE} with \(|\omega_i-\omega_j|< \max\{\Gamma_{ij},\Lambda_{ij}\}\) satisfies \(\Gamma_{ij}(\omega_i)\simeq \Gamma_{ij}(\omega_j)\simeq\tilde{\Gamma}_{ij}\) and \(\Lambda_{ij}(\omega_i)\simeq \Lambda_{ij}(\omega_j)\simeq\tilde{\Lambda}_{ij}\), and the replacement is trivially precise. Conversely, when \(|\omega_i-\omega_j| > \max\{\Gamma_{ij},\Lambda_{ij}\}\) the accuracy of the replacement decreases, but the effect of these terms becomes negligible. Consequently, the preciseness of the replacement is irrelevant, and these terms can even be discarded through secularization.

After the replacement, \autoref{eq.mccauley} is an LME if (i) the energy correction, \(\Delta = \sum_{ij}\tilde{\Lambda}_{ij}\sigma_i^\dagger\sigma_j\), is Hermitian and (ii) the Kossakowski matrix, \(\tilde{\Gamma}_{ij}\), is positive semidefinite. In the examples from~\cite{mccauley_accurate-master_2020} both conditions were automatically satisfied. As for the system explored in~\cite{dfpv_vacuum-field-induced_2023}, the non-Hermitian terms were removed through an accurate secularization, and the geometric symmetry of the setup implied a positive semidefinite Kossakowski matrix. In general, however, these requirements are not guaranteed and an LME is not obtained. In the following, we elaborate on strategies to address both problems.

\section{Non-Hermiticity} \label{sec.nonHermiticity}

Here, we discuss how non-Hermiticity can appear in the energy correction Hamiltonian and propose two solutions that ensure Hermiticity.

\subsection{Description of the problem}\label{subsec.nonHerm-conditions}


As mentioned above, for the master equation to be an LME, \(\Delta = \sum_{ij}\tilde{\Lambda}_{ij}\sigma_i^\dagger\sigma_j\) must be Hermitian, which is automatically fulfilled when \(\tilde{\Lambda}\) is Hermitian. In principle, it is possible for \(\Delta\) to be Hermitian even when \(\tilde{\Lambda}\) is not, but this would require fortuitous cancellations between system and environment properties that do not occur in general. Thus, we focus on approaches to convert \(\tilde{\Lambda}\) into a Hermitian matrix.

To illustrate the origin of non-Hermitian terms in \(\Delta\), we consider the same three-level system coupled to a Lorentzian bath described before (cf. \autoref{fig.system}). Despite its simplicity, this model is enough to reveal the source of non-Hermitian terms in \(\Delta\). Simultaneously, the corresponding Kossakowski matrix is positive semidefinite, due to the absence of environment cross-correlations, allowing us to isolate the non-Hermiticity. According to \autoref{eq.mccauley}, the energy shift for this system is
\begin{equation} \label{eq.cpshift_geometric}
    {\Delta} =
    \begin{pmatrix}
        \tilde{\Lambda}_{1'1'} + \tilde{\Lambda}_{2'2'} & 0 & 0 \\
        0 & \tilde{\Lambda}_{11} & \tilde{\Lambda}_{12} \\
        0 & \tilde{\Lambda}_{21} & \tilde{\Lambda}_{22}
    \end{pmatrix},
\end{equation}
where \(\tilde{\Lambda}_{12} = \tilde{\Lambda}_{21} = \sqrt{\lambda(\omega_1)}\sqrt{\lambda(\omega_2)}\). Notice that \(\lambda(\omega)\) changes sign at \(\omega_M\). Thus, if \(\omega_1\) and \(\omega_2\) lie on opposite sides of \(\omega_M\), then \(\tilde{\Lambda}_{12} = \tilde{\Lambda}_{21} = i |\tilde{\Lambda}_{12}|\), and \(\Delta\) is non-Hermitian. 

\begin{figure}[htbp]
    \centering
    \includegraphics[width=\columnwidth]{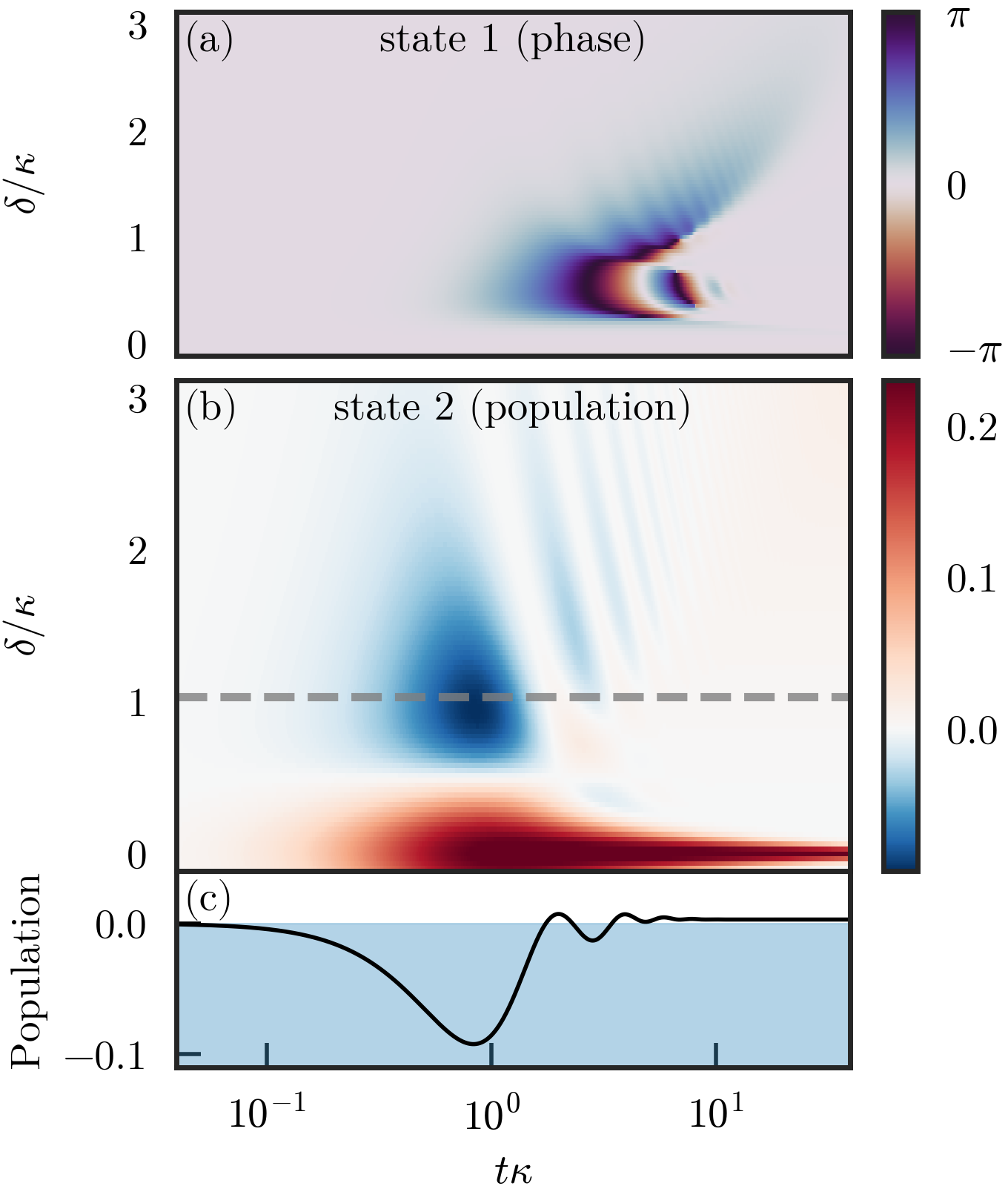}
    \subfloat{\label{subfig.geometric-11-map_phase}}
    \subfloat{\label{subfig.geometric-22-map_pop}}
    \subfloat{\label{subfig.geometric_22_cut}}
    \caption{Breakdown of the \(g\Lambda-g\Gamma\) prescription of~\cite{mccauley_accurate-master_2020}. (a) Non-trivial complex phase of the ``population'' of \(\ket{1}\) owing to the loss of Hermiticity of \autoref{eq.cpshift_geometric}. (b) Population of \(\ket{2}\) obtained with \autoref{eq.mccauley} with \(\omega_1 = \omega_M + \delta_1\), \(\omega_2 = \omega_M + \delta_2\) and \(\delta = -\delta_1 =\delta_2\) (very close to being real). Blue and red represent positive and negative values, respectively. (c) Population of \(\ket{2}\) at \(\delta = \kappa\). The blue region indicates negative, unphysical population. \label{fig.geometric_complex}}
\end{figure}

The situation described above illustrates how non-Hermiticity can arise in the energy correction operator. When this occurs, \autoref{eq.mccauley} is not truly an LME, and the positivity of \(\rho\) is no longer guaranteed. Not only that, but the Hermiticity of \(\rho\) is also not preserved and non-negligible imaginary contributions arise in the populations. As an example, we show in \autoref{fig.geometric_complex} the prediction of \autoref{eq.mccauley}, with parameters set to \(\hbar \omega_M = 1\)~eV, \(\hbar\kappa = 0.1\)~eV, \(g=\kappa\) and \(\omega_i=\omega_M+\delta_i\), choosing \(\delta_2 = -\delta_1 = \delta\), and \(\ket{\psi(0)} = \ket{1}\) as the initial state. In \autoref{subfig.geometric-11-map_phase}, we depict the phase of the ``complex population'' of \(\ket{1}\). While it starts at 0 for short time scales, it undergoes rapid changes and displays multiple complete turns. Thus, the Hermiticity of \(\rho\) is quickly compromised due to the off-diagonal, non-Hermitian couplings described above. In the middle and bottom panels, we plot the population of state \(\ket{2}\), showing only its real part, as the imaginary part is negligible. Nevertheless, \autoref{subfig.geometric-22-map_pop} shows positive and negative populations in red and blue, respectively, and the negative portions are definitely not negligible. In the bottom plot, a cut at \(\delta=\kappa\) accentuates the negativity, indicated with the blue region.

Despite the observations in \autoref{fig.geometric_complex}, note that these deficiencies do not necessarily mean that \autoref{eq.mccauley} with a non-Hermitian \(\Delta\) is always inaccurate. Similarly to the BRE, it can remain precise if the underlying approximations are reasonable~\cite{hartmann_embracing-nonpositivity_2020}. The parameters have indeed been chosen to highlight the difficulties that can arise from \autoref{eq.mccauley}. Still, it remains true that the attempt at a solution to the BRE problems has lead to a new, arguably worse issue: populations become complex rather than simply negative. In the following subsections, we describe and compare several alternatives designed to eliminate the non-Hermiticity problem.

\subsection{Approach 1: Degenerate levels} \label{subsec.degenerate}

It was argued in the introduction that the off-diagonal terms are relevant when the energetic difference between states \(\ket{1}\) and \(\ket{2}\) in \autoref{fig.system} is comparable to the energy corrections. For that reason, a possible solution to the Hermiticity problem would be to ignore the energy difference between these states while deriving the master equation, and then add it back as a correction afterwards. In the illustrative system of \autoref{fig.system}, this corresponds to computing the parameters in the master equation as if \(\omega_1=\omega_2=\bar{\omega} = (\omega_1 + \omega_2)/2\), and then adding the actual energy difference to the effective Hamiltonian. We follow up on the notation introduced previously and refer to this method as \(d\Lambda-d\Gamma\), as the close-lying energy levels of the system are assumed to be degenerate from the start in the master equation derivation. With this procedure, the energy shift is
\begin{equation} \label{eq.cpshift_degenerate}
    \Delta^d =
    \begin{pmatrix}
        2\lambda(-\bar{\omega}) & 0 & 0 \\
        0 & \lambda(\bar{\omega}) & \lambda(\bar{\omega}) \\
        0 & \lambda(\bar{\omega}) & \lambda(\bar{\omega})
    \end{pmatrix},
\end{equation}
and, since \(\lambda(\bar{\omega})\) is real, \(\Delta^d\) is automatically Hermitian. This replacement is numerically accurate when \(J(\omega_2) \) and \(J(\omega_1) \simeq J(\bar{\omega}) \). However, we do not expect it to perform very well whenever either \(J(\omega_2)\) or \(J(\omega_1)\not\simeq J(\bar{\omega})\) because, although \(\Delta^d_{12}\) becomes negligible, the dissipative terms are also evaluated at \(\bar{\omega}\), significantly different from \(\omega_1\) and \(\omega_2\), potentially leading to inaccurate decay rates.

This procedure is easily applicable to systems exhibiting clearly structured energy spectra with several narrow sets of levels, because one single frequency can be reliably assigned to each set. Within each set, the off-diagonal couplings would resemble the Hermitian expression from \autoref{eq.cpshift_degenerate}, and although the couplings between different sets would still be susceptible to the non-Hermiticity issue, these can in principle be eliminated through secularization. However, for more complicated level structures, where the energy levels are not distinctly arranged in narrow sets, assigning the same frequency to several states becomes a less straightforward task. Exquisite care must then be taken in order to avoid losing pertinent information about the frequency dependence of \(J\) and \(\lambda\), which impacts the precise values of the energy shifts and decay rates. Additionally, this means that in general the procedure has to be performed manually, or at least under strict human supervision.

\subsection{Approach 2: Arithmetic mean} \label{subsec.arithmetic}

We devise a second approach for addressing the non-Hermitian couplings that involves replacing the geometric mean discussed in section \ref{subsec.mccauley} with the arithmetic mean solely in the energy correction, while keeping the geometric mean in the decay terms:
\begin{subequations}
\begin{align}
    \tilde{\Lambda}_{ij} &= \frac{\Lambda_{ij}(\omega_i) + \Lambda_{ij}(\omega_j)}{2},\\
    \tilde{\Gamma}_{ij} &= \sqrt{\Gamma_{ij}(\omega_i)}\sqrt{\Gamma_{ij}(\omega_j)}.
\end{align}
\end{subequations}
Accordingly, we designate this prescription as \(a\Lambda-g\Gamma\). This adjustment allows for the refactoring of the energy shift terms in \autoref{eq.BRE} into a commutator. It is worth noting that if the arithmetic mean were used in the decay terms as well, the Kossakowski matrix for the system in \autoref{fig.system} would give rise to a negative decay rate, approximately given by 
\begin{equation}    
    \pi\left(J_+ - \sqrt{J_+^2 + J_-^2}\right) < 0,
\end{equation}
where \(J_\pm = J(\omega_1)\pm J(\omega_2)\). Similar features appear also in other, more complex systems.

With this option, the energy correction for the three-level system becomes 
\begin{equation}\label{eq.cpshift_arithmetic}
    \Delta^a =
    \begin{pmatrix}
        \lambda(-\omega_1)+\lambda(-\omega_2) & 0 & 0 \\
        0 & \lambda(\omega_1) & \frac{\lambda(\omega_1)+\lambda(\omega_2)}{2} \\
        0 & \frac{\lambda(\omega_1)+\lambda(\omega_2)}{2} & \lambda(\omega_2)
    \end{pmatrix}.
\end{equation}
Unlike the previous fix, the arithmetic mean naturally preserves the information about the dependence of \(J\) and \(\lambda\) with \(\omega\). Hence, this solution to the non-Hermiticity problem is more straightforwardly applicable to more complex systems. We finish by noting that this approach yields an expression for \(\Delta\) similar to the equivalent term in the second line of \autoref{eq.BRE2}, an alternative form of the BRE. However, in that equation, the \(\Lambda_{ij}\) functions appear together with an additional \(i(\Gamma_{ij}(\omega_i)-\Gamma_{ij}(\omega_j))/4\).

\subsection{Accuracy comparison} \label{subsec.comparison_hermiticity}

\begin{figure*}[htbp]
    \centering
    \subfloat{\label{subfig.rho22_exact}}
    \subfloat{\label{subfig.rho22_arith}}
    \subfloat{\label{subfig.rho22_deg}}
    \subfloat{\label{subfig.rho22_deg2}}
    \includegraphics[width=\textwidth]{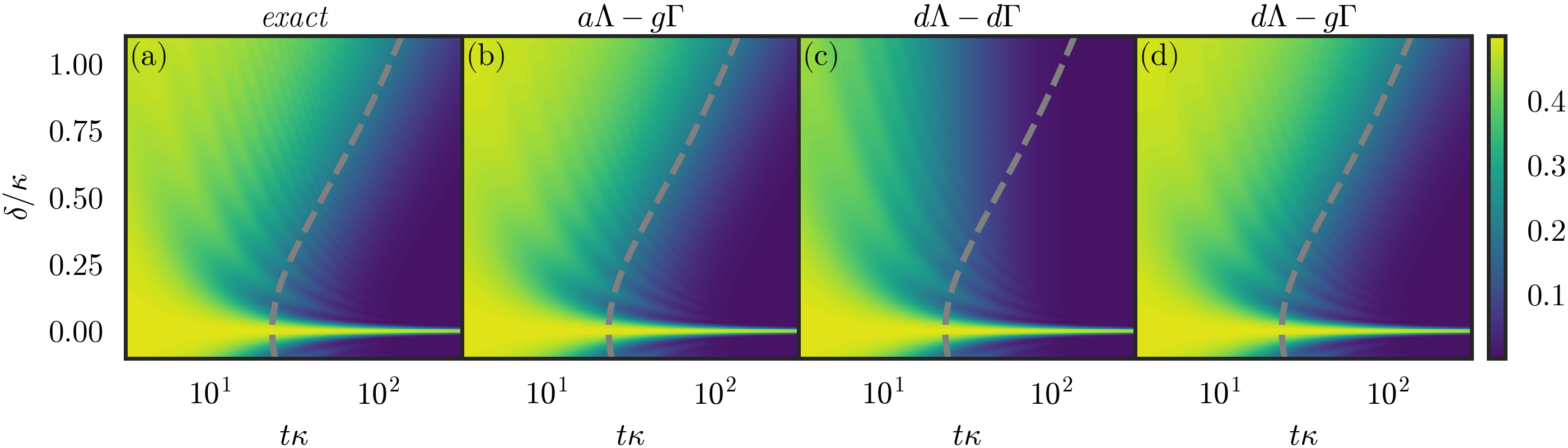}
    \caption{Population of the state \(\ket{2}\) calculated with (a) the \textit{exact}, (b) \(a\Lambda-g\Gamma\) and (c) \(d\Lambda-d\Gamma\) methods, as a function of \(\delta\). (d) \(d\Lambda-nd\Lambda\), designed to fix the \(d\Lambda-d\Gamma\) method by keeping the full information of the system frequencies. The gray dashed line is proportional to \(1/J(\omega_M+\delta)\), indicating the lifetime of \(\ket{2}\) when the excited states do not interfere. The parameters are \(\hbar\omega_M = 1\)~eV, \(\hbar \kappa = 0.1\)~eV, \(g=0.05\kappa\), \(\omega_1=\omega_M+\delta_1\) and \(\omega_2=\omega_M+\delta_2\), with \(-\delta_1 = \delta_2=\delta\). The initial state is \(\ket{\psi(0)} = (\ket{1}-\ket{2})/\sqrt{2}\).
    \label{fig.pop_arith-deg}}
\end{figure*}

In this section, we evaluate the dynamics of the three-level system depicted in \autoref{fig.system} using several methods. We then compare and identify the best option among the approaches from sections \ref{subsec.degenerate} and \ref{subsec.arithmetic}, referred to as \(d\Lambda-d\Gamma\) and \(a\Lambda-g\Gamma\), respectively. Additionally, we include an \textit{exact} method that exploits the well-known equivalence between a Lorentzian bath and a single bosonic mode, \(a\), coupled to a flat (Markovian) environment, from the perspective of the system~\cite{imamoglu_non-markovian_1994}. Therefore, the system plus mode system can be exactly described through
\begin{subequations} 
    \begin{align}  \nonumber
        \dot{\rho} = &- \frac{i}{\hbar}\left[H + g d\left(a+a^\dagger\right) + \hbar\omega_M a^\dagger a, \rho\right] \\ \label{eq.modeME}
        &+ \frac{\kappa}{2}\left(-\left\{a^\dagger a, \rho\right\} + 2 a\rho a^\dagger\right), \\
       d = &\sum_{i=1}^2 \left(\sigma_i + \sigma_i^\dagger\right),
    \end{align}
\end{subequations}
where the parameters are defined as in \autoref{eq.notation1} and \autoref{eq.notation2}, and the system bath Hamiltonian from \autoref{eq.systembath} has been replaced with the simpler coupling to a single mode. From \autoref{eq.modeME}, we may obtain the reduced system density matrix by tracing over the bosonic degree of freedom. The approximate methods can then be compared to the \textit{exact} solution.

For the comparison, we choose the following parameters: \(\hbar\omega_M = 1\)~eV, \(\hbar \kappa = 0.1\)~eV, \(g=0.05\kappa\), \(\omega_1=\omega_M+\delta_1\) and \(\omega_2=\omega_M+\delta_2\), with \(-\delta_1 = \delta_2=\delta\). The initial state is \(\ket{\psi(0)} = (\ket{1}-\ket{2})/\sqrt{2}\). We show the population of state \(\ket{2}\) as calculated with the \textit{exact}, \(a\Lambda-g\Gamma\) and \(d\Lambda-d\Gamma\) methods in \autoref{subfig.rho22_exact}, \autoref{subfig.rho22_arith} and \autoref{subfig.rho22_deg}, respectively. When the states \(\ket{1}\) and \(\ket{2}\) are close to degeneracy (\(\delta/\kappa \approx 0\)), both the \(a\Lambda-g\Gamma\) and the \(d\Lambda-d\Gamma\) methods agree with the \textit{exact} result, as expected. However, as \(\delta/\kappa\) increases, the \textit{exact} and \(a\Lambda-g\Gamma\) color maps reveal a gradual and clear increase in the decay timescale, marked with dashed lines for ease of comparison between plots. In contrast, this effect is not present in the \(d\Lambda-d\Gamma\) method. 

It should be noted that the reason for the disparity of the \(d\Lambda-d\Gamma\) method compared to the first two lies in the decay terms, rather than in \(\Delta\). This is because the \(d\Lambda-d\Gamma\) implementation approximates the levels as degenerate during the full derivation of the master equation, and only afterwards includes the finer structures. Hence, both \(\lambda\) and \(J\) are evaluated at \(\bar{\omega}\) and, with the parameters used above, the decay rate far from resonance is still given by \(2\pi J(\omega_M)\) in \autoref{subfig.rho22_deg}, while it becomes smaller for increasing values of \(\delta/\kappa\) in \autoref{subfig.rho22_exact} and \autoref{subfig.rho22_arith}. This can be corrected by not replacing \(\omega_1\) and \(\omega_2\) with \(\bar{\omega}\) in the decay part, as shown in \autoref{subfig.rho22_deg2}, where the same dependence of the system's lifetime on \(\delta/\kappa\) is observed. With our notation, we can term this ``mixed approach'' \(d\Lambda-g\Gamma\). Judging from \autoref{subfig.rho22_deg2}, we conclude that, in the situation considered here, both \(a\Lambda-g\Gamma\) and \(d\Lambda-g\Gamma\) work well. Still, the energy shift in \(d\Lambda-g\Gamma\) assumes no natural detuning between the excited states, while in reality such frequency difference is non-zero. Therefore, it could be the case that some concrete scenarios might compromise the accuracy of \(d\Lambda-g\Gamma\) compared to \(a\Lambda-g\Gamma\), where the detuning between excited states is automatically included with no additional consideration, thus ensuring a more robust description of the emitter's dynamics.

Considering these findings, we conclude that the \(a\Lambda-g\Gamma\) approach from section \ref{subsec.arithmetic} provides the best solution. It successfully reproduces the exact solution in the weak coupling regime, addresses the non-Hermiticity issue associated with the geometric mean prescription, and seamlessly incorporates the energetic features of the system without additional supervision. This makes the \(a\Lambda-g\Gamma\) approach a robust and effective method for describing the energy shift of complex quantum systems. We have also tried several other options, such as manually ensuring that \(\tilde{\Lambda}_{ij} = \tilde{\Lambda}_{ji}^*\) by arbitrarily conjugating one of them. For conciseness, these methods are not discussed here, as they are worse and more arbitrary than the arithmetic mean procedure.

\section{Non-positive semidefiniteness} \label{sec.nonpositivity}

We continue our analysis by clarifying the situations that ensure the positive semidefiniteness of the Kossakowski matrix. Notably, in the case of a single system operator \(A\) coupling to one bath operator \(B\), such that \(H_\mathrm{sb}=AB\), the ensuing Kossakowski matrix is positive semidefinite by construction, as demonstrated in~\cite{mccauley_accurate-master_2020}. Indeed, the procedure outlined in \autoref{subsec.mccauley} yields a dissipative term of the form \(\frac12 \left(-\left\{\Sigma^\dag\Sigma, \rho\right\} + 2 \Sigma\rho\Sigma^\dag \right)\), with a single ``collective'' decay operator \(\Sigma = \sum_j A_j \sqrt{\gamma(\omega_j)}\sigma_j\). Then, the only non-zero eigenvalue of the Kossakowski matrix is a positive 1.

Nevertheless, the scenario becomes more intricate when allowing for multiple system-bath interaction terms, \(H_\mathrm{sb} = \sum_\alpha  A_\alpha  B_\alpha \), with the potential appearance of bath cross-correlations, i.e., off-diagonal elements in \autoref{eq.gamma}. In this case, the Kossakowski matrix after the geometric mean replacement can often be non-positive semidefinite. Yet, specific conditions can guarantee non-negative eigenvalues. For instance, if the cross-correlations vanish and each system transition involves only one of the \(A_\alpha\) operators, the situation reduces to a sum over several single system-bath interactions. These requirements can only be met in very particular configurations, like the one considered in~\cite{dfpv_vacuum-field-induced_2023}, which can be a significant limitation.

In the following, we perform a structured study of the size and impact of the negative eigenvalues of the Kossakowski matrix. First, we focus our analysis on a specific system of small size, and observe that, with the prescription \(a\Lambda-g\Gamma\) from earlier, the negative eigenvalues can be discarded without significant accuracy loss. Next, we examine a large sample of randomly generated test systems to confirm the previous conclusion that the negative eigenvalues can be neglected. At the moment, we do not have a rigorous justification as to why taking the geometric mean should imply that, but the statistical study seems to confirm it to be the case.

\subsection{Particular case} \label{subsec.particular}

Here, we generate a particular system and explore the effectiveness of various approaches aimed at achieving the positivity of the corresponding Kossakowski matrix. The system is generated according to the following procedure. We consider a generic \(L\)-level system coupled to the environment through \(M\) interaction terms, such that \(H_\mathrm{sb} = \sum_{\alpha =1}^M A_\alpha  B_\alpha \). The spectral density of the bath is, then, represented as a positive semidefinite \(M\times M\) matrix-valued function. To describe the properties of the environment, we exploit the equivalence between a generic bath and a particular set of \(N\) coupled and lossy modes \(a_\beta\), as discussed in~\cite{medina_fewmodes_2021, barquilla_fewmode-multiemitter_2022}. This methodology is a generalization of the one employed in \autoref{subsec.comparison_hermiticity}, where a single discrete and lossy mode was enough to capture the full effect of a Lorentzian bath on the emitter, and the simple \autoref{eq.modeME} yielded the exact dynamics. By allowing the modes to increase in number and be coupled to each other, more complex spectral densities can be represented with just a few discrete modes. It can be shown, then, that the associated spectral density is
\begin{subequations}
    \begin{equation} \label{eq.J_fit}
        \mathbf{J}(\omega) = \frac\hbar\pi \mathbf{g} \cdot \textrm{Im}\left\{(\mathbf{h} - \omega)^{-1}\right\} \cdot \mathbf{g}^T,
    \end{equation}
    where particular matrices \(\mathbf{h}\) (\(N\times N\)) and \(\mathbf{g}\) (\(M\times N\)), along with an integer \(N\), can be chosen to accurately fit any spectral density. The corresponding Lamb shift integral is given by 
    \begin{equation} \label{eq.L_fit}
        \boldsymbol{\lambda}(\omega) = - \hbar\mathbf{g} \cdot \textrm{Re}\left\{(\mathbf{h}-\omega)^{-1}\right\} \cdot \mathbf{g}^T.
    \end{equation}
\end{subequations}
More specifically, \(h_{\beta\beta}=\Omega_{\beta\beta} - i \kappa_\beta/2\) represents the frequencies and losses of the modes, \(h_{\beta\gamma} = \Omega_{\beta\gamma}\) contains the couplings between discrete modes, and \(g_{\alpha\beta}\) determines the coupling strength between the mode \(a_\beta\) and the system through \(A_\alpha\). Additionally, using the results from~\cite{barquilla_fewmode-multiemitter_2022}, the exact dynamics of the system is also available through
\begin{subequations} \label{eq.exactDynamics}
    \begin{align}\nonumber
        \dot{\rho}_\mathrm{sm} ={}&-\frac{i}{\hbar}\left[ H_\mathrm{sm}, \rho_\mathrm{sm} \right] \\
        {}&+ \sum_{\beta=1}^{N} \frac{\kappa_\beta}{2} \left(-\left\{a_\beta^\dag a_\beta, \rho_\mathrm{sm}\right\} + 2 a_\beta\rho_\mathrm{sm}a_\beta^\dag\right), \\ \nonumber
        H_\mathrm{sm} ={}& H + \sum_{\beta\gamma=1}^N \hbar\Omega_{\beta\gamma}a_\beta^\dag a_\gamma \\ 
        & + \sum_{\alpha\beta=1}^{M,N} \hbar A_\alpha   g_{\alpha \beta}\left(a_\beta^\dag + a_\beta \right),\\ 
        \rho ={}& \mathrm{Tr}_\mathrm{m} \{\rho_\mathrm{sm}\},
    \end{align}
    where \(\rho_\mathrm{sm}\) is the density matrix of the system and the discrete modes.
\end{subequations}
Therefore, \autoref{eq.exactDynamics} provides a natural benchmark for any perturbative equation we propose. 

\begin{figure}[htbp]
    \centering
    \subfloat{\label{subfig.particular_system}}
    \subfloat{\label{subfig.particular_spectral}}
    \subfloat{\label{subfig.particular_dynamics}}
    \subfloat{\label{subfig.particular_error}}
    \includegraphics[width=\columnwidth]{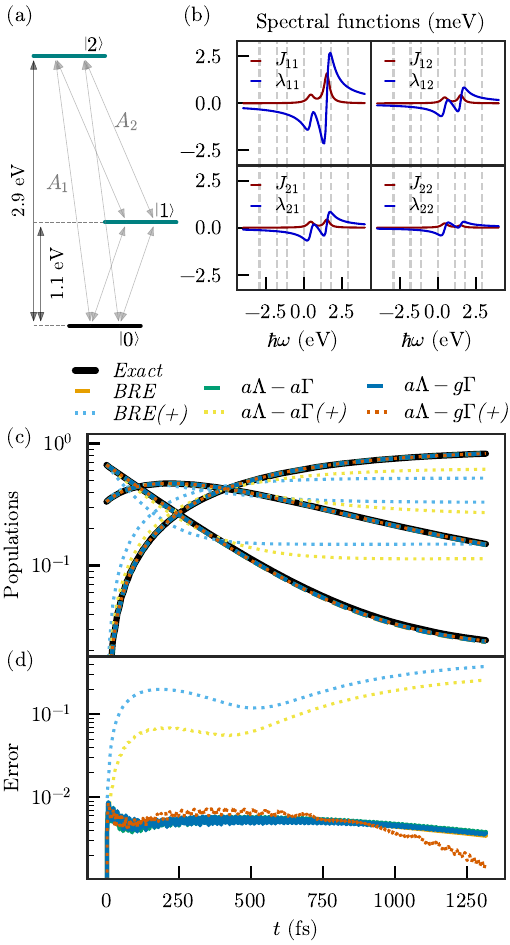}
    \caption{Particular case: (a) Scheme of the levels and interaction operators. (b) Spectral density (red) and Lamb shift integral (blue). The faint dashed lines represent the 6 different transition energies present in the system. (c) Time evolution of the populations. (d) Time evolution of the deviation with respect to the exact dynamics. The dashed lines have been calculated with the ``full'' methods including negative and positive eigenvalues, and the dotted lines with their ``positive'' counterparts. \label{fig.particular_system}}
\end{figure}

For the analysis of the specific system, we set the number of levels, interaction terms and discrete modes to \(L=3\), \(M=2\) and \(N=2\); with any simpler choice, the geometric mean (\(g\Gamma\)) yields a positive semidefinite Kossakowski matrix. In \autoref{subfig.particular_system} and \autoref{subfig.particular_spectral}, we schematically show the energy structure of the system and the spectral functions of the environment, produced with the method described above with random parameters. The vertical dashed lines in \autoref{subfig.particular_spectral} represent all the possible transition energies in the system. We then solve the dynamics of the particular system with \(\ket{\psi(0)} = (\ket{1} + \sqrt{2}\ket{2})\sqrt{3}\) as the initial state. To that end, we employ seven methods, namely, \textit{Exact}, \textit{BRE}, \textit{BRE}(+), \(a\Lambda-a\Gamma\), \(a\Lambda-a\Gamma\)(+), \(a\Lambda-g\Gamma\) and \(a\Lambda-g\Gamma\)(+). The \textit{Exact} and \textit{BRE} methods solve \autoref{eq.exactDynamics} and \autoref{eq.BRE}, respectively. As for the \(a\Lambda-a\Gamma\) and \(a\Lambda-g\Gamma\) methods, we take our own advice from \autoref{subsec.comparison_hermiticity} and represent the energy shift always following the arithmetic mean prescription: \(\tilde{\Lambda}_{ij} = (\Lambda_{ij}(\omega_j) + \Lambda_{ij}(\omega_i))/2\). The various approaches differ in the procedure to obtain the Kossakowski matrix, i.e., the dissipative part. For the \(a\Lambda-a\Gamma\) we take the arithmetic mean, \(\tilde{\Gamma}_{ij} = (\Gamma_{ij}(\omega_j) + \Gamma_{ij}(\omega_i))/2\), and for the \(a\Lambda-g\Gamma\) the geometric mean, \(\tilde{\Gamma}_{ij} = \sqrt{\Gamma_{ij}(\omega_j)}\sqrt{\Gamma_{ij}(\omega_i)}\). Last, all 3 approximate methods have a positive, (+), counterpart, in which the Kossakowski matrix is diagonalized, and the negative eigenvalues are discarded. As a consequence, the (+) methods are all genuine LMEs, and the question is which of these, if any, accurately represents the dynamics of the system.

In \autoref{subfig.particular_dynamics} and \autoref{subfig.particular_error}, we display the evolution of the populations as obtained from the methods discussed above and the deviation with respect to the exact result, respectively. The deviation is obtained through the Frobenius norm of the corresponding density matrices: \(||\rho_\mathrm{Exact}(t) - \rho_\mathrm{Method}(t)||_F\). First, note that every ``full'' method including both positive and negative eigenvalues of the Kossakowski matrix agrees with the exact result, as expected in a sufficiently weak coupling regime. Accordingly, the dashed lines lie all on top of each other. Regarding the positive versions of the methods, there are two sets of lines distinctly separate from the rest, concretely, the ones corresponding to the \textit{BRE}(+) and \(a\Lambda-a\Gamma\)(+). It can be therefore concluded that removing the negative eigenvalues of the Kossakowski matrix yields incorrect predictions for the BRE and the arithmetic mean. However, rather remarkably, there is barely any difference between the \(a\Lambda-g\Gamma\) and \(a\Lambda-g\Gamma\)(+) methods. Thus, at least in this particular case, it seems that the geometric mean greatly reduces the effect of the negative eigenvalues of the Kossakowski matrix. To be more explicit, we diagonalize the Kossakowski matrices corresponding to the \textit{BRE}, \(a\Lambda-a\Gamma\) and \(a\Lambda-g\Gamma\) methods and present the most significant eigenvalues in ascending order (first two and last two) in \autoref{tab.eigenvalues}. There, the blue and red columns contain the two largest negative and positive eigenvalues, respectively. We observe a striking difference in the ratio between the largest negative and positive eigenvalues, which is \(\mathcal{O}(-1)\) for both the \textit{BRE} and \(a\Lambda-a\Gamma\) methods, but only \(\mathcal{O}(-10^{-2})\) for the \(a\Lambda-g\Gamma\) approach. The smallness of the negative eigenvalues after performing the geometric mean procedure in the dissipative part explains why their removal does not imply a loss of accuracy from \(a\Lambda-g\Gamma\) to \(a\Lambda-g\Gamma\)(+).
\begin{table}[hbtp]
    \centering
    \caption{Negative (blue) and positive (red) eigenvalues of the corresponding Kossakowski matrices. We show only the 4 largest in absolute value, out of the 9 total eigenvalues for \(L=3\), rounded to the first significant digit.}
    \label{tab.eigenvalues}
    \begin{tblr}{
        colspec={cccccc},
    }
        \toprule
        \textbf{Method} & \SetCell[c=5]{c} \textbf{Eigenvalues} (\(\mathrm{meV}/\hbar\)) \\
                        & 1st & 2nd & \dots & 8th & 9th \\
        \midrule 
        \textit{BRE} & {\color{MidnightBlue}\(-5\)} &{\color{MidnightBlue} \(-9\cdot10^{-2}\)}  & & {\color{BrickRed} \(2\cdot 10^{-1}\)} & {\color{BrickRed} \(9\)} \\  
        \(a\Lambda-a\Gamma\) & {\color{MidnightBlue}\(-2\)} & {\color{MidnightBlue} \(-5\cdot10^{-2}\)} & & {\color{BrickRed} \(9\cdot10^{-3}\)} & {\color{BrickRed} \(6\)}\\
        \(a\Lambda-g\Gamma\) & {\color{MidnightBlue} \(-5\cdot10^{-2}\) }& {\color{MidnightBlue} \(-5\cdot10^{-5}\)} & & {\color{BrickRed} \(1\cdot10^{-2}\)} & {\color{BrickRed} \(4\)} \\
        \bottomrule
    \end{tblr}
\end{table}

\subsection{Statistical study} \label{subsec.statistical}

\begin{figure}[htbp]
    \centering
    \subfloat{\label{subfig.hist_bre}}
    \subfloat{\label{subfig.hist_diffArith}}
    \subfloat{\label{subfig.hist_diffGeo}}
    \includegraphics[width=\columnwidth]{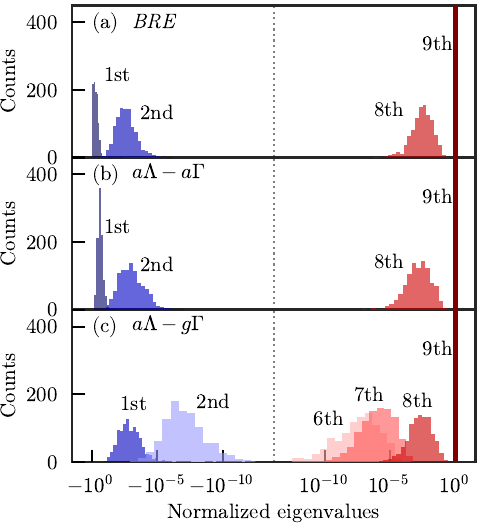}
    \caption{Distribution of the most significant eigenvalues of the Kossakowski matrix, obtained with the (a) \textit{BRE}, (b) \(a\Lambda-a\Gamma\) and (c) \(a\Lambda-g\Gamma\) methods. Blue and red histograms represent negative and positive eigenvalues, respectively. Each eigenvalue is specified by the shade of the color. The horizontal axes are split along the middle of each plot (vertical dotted line), and display a negative or positive logarithmic scale.}
    \label{fig.hist}
\end{figure}
In this subsection, we check whether the reduction of the negative eigenvalues after the geometric mean procedure in the Kossakowski matrix generalizes in a random sample of systems. To that end, we again exploit \autoref{eq.J_fit} and \autoref{eq.L_fit} to generate a sample of 1000 systems and baths with random parameters. More details can be found in Appendix \ref{app.random}. Then, to assess the consistency of the size reduction of the negative eigenvalues following the geometric mean procedure, we diagonalize the resulting Kossakowski matrices and plot the eigenvalue distributions in \autoref{fig.hist}, where blue and red histograms indicate negative and positive eigenvalues, respectively. To homogenize the systems, we normalize the eigenvalues to the largest positive one (9th) in these plots, hence the dark red peak at \(10^0\). Notice that the dark blue histograms close to \(-10^{0}\) in \autoref{subfig.hist_bre} and \autoref{subfig.hist_diffArith} indicate the presence of large negative eigenvalues for the \textit{BRE} and \(a\Lambda-a\Gamma\) methods. However, rather remarkably, the histogram corresponding to the first eigenvalue in \autoref{subfig.hist_diffGeo} does not cluster close to \(-10^{0}\), but around \(-10^{-3}\). This is in agreement with the previous observation that the geometric mean significantly reduces the negative eigenvalues. 
Therefore, \autoref{fig.hist} supports the conclusion that removing the negative eigenvalues after the geometric mean procedure, and thus reaching an LME, should not impact the accuracy of the QME. Incidentally, we also remark here that the removal of the negative eigenvalues is known to yield the closest symmetric positive semidefinite matrix~\cite{higham_nearestpositive_1988}. 

\begin{figure}[htbp]
    \centering
    \subfloat{\label{subfig.dev_bre}}
    \subfloat{\label{subfig.dev_diffArith}}
    \subfloat{\label{subfig.dev_diffGeo}}
    \subfloat{\label{subfig.dev_breP}}
    \subfloat{\label{subfig.dev_diffArithP}}
    \subfloat{\label{subfig.dev_diffGeoP}}
    \includegraphics[width=\columnwidth]{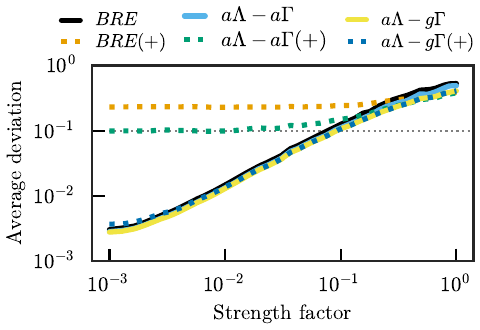}
    \caption{Statistical average over the random systems of the time-averaged deviation of the dynamics (\autoref{eq.average_deviation}), as a function of the coupling strength factor. The continuous lines refer to the ``full'' methods, keeping the negative eigenvalues, and the dotted lines represent the ``positive'' methods. The horizontal dotted line indicates the point beyond which any method undoubtedly fails to reproduce the exact dynamics.
    \label{fig.deviations}}
\end{figure}

In order to gain information relative to the breakdown of the equations due to the onset of the strong coupling regime, we introduce a numerical factor in the spectral functions to artificially scale up the interaction for each random system instance, spanning a range of several orders of magnitude. We then solve the dynamics with each method and evaluate the deviation from the exact evolution in the whole sample. For the sake of thoroughly exploring the parameter space, we also randomly choose the initial state \(\ket{\psi(0)}\) for each system instance. The deviation is measured with the time-averaged Frobenius norm of the difference between the density matrices:
\begin{equation} \label{eq.average_deviation}
    \Delta_\mathrm{Method} = \frac1T \int_{0}^{T} \mathrm{d} t\ || \rho_\mathrm{Exact}(t) - \rho_\mathrm{Method}(t)||_F.
\end{equation}
In practice, however, it is more sensible to study the statistical properties of \(\log\Delta_\mathrm{Method}\), to account for the skewness of \(\Delta_\mathrm{Method}\), which is due to \(\Delta_\mathrm{Method}\) being small but always positive. The results are condensed in \autoref{fig.deviations}, where the average deviation is \(\exp({\langle\log\Delta_\mathrm{Method}\rangle})\) or, equivalently, the geometric mean. The first main piece of information presented there is that the ``full'' methods (continuous lines) all converge to the exact result when the coupling strength becomes smaller, as expected. We show with the dotted lines that removing the negative eigenvalues of the Kossakowski matrix from the \textit{BRE} and \(a\Lambda-a\Gamma\) approaches results in the stabilization of the average deviation around 0.1, rather than convergence to the exact solution. Hence, we observe that the importance of the negative eigenvalues is not really dependent on the coupling strength in those two methods. Strikingly, notice that the dotted line corresponding to the \(a\Lambda-g\Gamma\)(+) method follows closely the deviation of the ``full'' methods. Therefore, the effect of the negative eigenvalues is small enough in the \(a\Lambda-g\Gamma\) method that they can be safely neglected, a procedure that automatically yields a genuine LME. The lines in \autoref{fig.deviations} imply that we have found an approximate LME that works accurately in the same coupling regime as the BRE, with three big advantages. First, our \(a\Lambda-g\Gamma\)(+) master equation is always free from unphysical effects. Second, because the \(a\Lambda-g\Gamma\)(+) master equation is a true LME, it is a completely positive map, which means that its lifted dynamics is positive for any additional subsystem too. In this regard, a reasonable question is whether the lifted dynamics can be used to gain more information about the accuracy of various LME-restoring approaches. Doing this comparison, we find that there is no significant difference between simulating a simple system or an extended one, with respect to the average deviation. We therefore do not show these results explicitly. The last advantage is that the \(a\Lambda-g\Gamma\)(+) QME admits a quantum trajectory interpretation of the dynamics with non-negative jump probabilities. 

We end the discussion with two last comments regarding the statistical distribution of the deviations. First, from our large sample size we can calculate the standard deviation of the distribution of \(\log\Delta_\mathrm{Method}\). Doing so yields a standard deviation of approximately a quarter of an order of magnitude quite consistently across different methods and coupling strengths. This means that the lines in \autoref{fig.deviations} have an approximate width of half an order of magnitude. We have omitted the dispersion measurement in \autoref{fig.deviations}, as it did not provide much insight and only made the figure harder to interpret. Secondly, from \autoref{fig.deviations} it may seem as though all approximate master equations break down equally when the strength factor grows. This is, however, not exactly true. Taking the logarithm before averaging the data camouflages the fact that, in some of the 1000 systems, the \textit{BRE} and \(a\Lambda-a\Gamma\) methods return divergent populations. The number of these pathological systems increases with the strength factor, and is linked to the non-positivity of their Kossakowski matrices. In that sense, the \textit{BRE} and \(a\Lambda-a\Gamma\) methods can suffer from a worse breakdown than the other methods. Although the \(a\Lambda-g\Gamma\) approach could have the same problem, in the studied coupling regimes it does not, in line with the smallness of the negative eigenvalues of its Kossakowski matrix. Of course, true LMEs such as the one proposed here, \(a\Lambda-g\Gamma\)(+), have a positive semidefinite Kossakowski matrix and are, therefore, completely free from these issues.

\section{Conclusion} \label{sec.conclusion}

Our work shows that the methodology developed in~\cite{mccauley_accurate-master_2020} and extended in~\cite{dfpv_vacuum-field-induced_2023} to derive an LME was not free of limitations, which prevented the final equation from being a desired LME in certain regimes. In particular, the Lamb shift Hamiltonian was susceptible to non-Hermiticity, leading to the loss of the Hermiticity of the density matrix and, accordingly, to complex populations. Additionally, the Kossakowski matrix associated with the derived master equation exhibited non-positive semidefiniteness, violating one of the assumptions of Lindblad's theorem, except under relatively simple configurations. We solve the Hermiticity concern by replacing the geometric mean introduced in \autoref{subsec.mccauley} with an arithmetic mean in the energy shift terms. Among various alternatives, we highlight this approach due to its superior simplicity in implementation. Regarding the non-positive definiteness of the Kossakowski matrix, we observe that while the geometric mean replacement did not eliminate the negative eigenvalues, these become substantially smaller than those in the original BRE, and their impact on the dynamics is markedly suppressed. Therefore, we adopt a pragmatic perspective and propose the removal of the negative eigenvalues to satisfy the assumptions of Lindblad's theorem, after the geometric mean replacement. This option is supported by the statistical study performed in this work, and particular simulations of larger systems not shown here for brevity. The combined implementation of the solutions to both problems yields an accurate and general LME that, akin to the BRE, maintains close ties with the physical environment, while simultaneously overcoming the mathematical deficiencies inherent to the BRE and the physical inadequacy of the secular approximation.

An implementation of the master equation developed here, \(a\Lambda-g\Gamma(+)\), is available at \url{https://github.com/diego-fpv/lindbladAG}.

\section*{Acknowledgments}
This work has been funded by the Spanish Ministry of Science, Innovation and Universities-Agencia Estatal de Investigación through the FPI contract No.~PRE2019-090589 as well as grants PID2021-125894NB-I00, EUR2023-143478, and CEX2018-000805-M (through the María de Maeztu program for Units of Excellence in R\&D). We also acknowledge financial support from the Proyecto Sinérgico CAM 2020 Y2020/TCS-6545 (NanoQuCo-CM) of the Community of Madrid.

\appendix

\section{Random system generation details}
\label{app.random}

The system and bath parameters were stochastically generated using the Python library NumPy (version 1.26.2). We employ an initial seed equal to 1234321 for the random number generator for reproducibility. For the initial state, we generate both an amplitude and a phase for each coefficient. To align the energy scales of the system and discrete modes, the system energies \(\hbar\omega_{\alpha}\) were confined to the interval \([0.1~\mathrm{eV}, 5.0~\mathrm{eV})\), and the mode energies \(\hbar\Omega_{\beta\beta}\) to \([0.3~\mathrm{eV}, 2.0~\mathrm{eV})\). The mode couplings \(\hbar\Omega_{\beta\gamma}\) are symmetric and ranged from \(0.0\)~eV to \(1.0\)~eV, and the rates \(\hbar\kappa_\beta\) were sampled from \([0.2~\mathrm{eV}, 0.5~\mathrm{eV})\). The transition operators were constructed as Hermitian matrices with zeros on the diagonal and values ranging from 0 to 1 elsewhere. Their physical dimensions are absorbed in the spectral functions, making the transition operators unitless. Concerning the couplings, \(\hbar g_{\alpha\beta}\) were uniformly distributed within the interval \([0.0~\mathrm{meV}, 1.0~\mathrm{meV})\), scaled by the square root of the strength factor discussed in \autoref{subsec.statistical}. This parametrization ensures a diverse set of systems for our analysis.

\bibliography{bibliografia}

\end{document}